\documentclass[runningheads, orivec]{llncs}
\usepackage[T1]{fontenc}
\usepackage{svg}
\usepackage{cite}
\usepackage{esvect}
\usepackage{booktabs}
\usepackage{hyperref}
\usepackage{graphicx}
\usepackage{amsmath}
\usepackage{comment}

\begin{document}
%
\title{Towards Trustworthy Breast Tumor Segmentation in Ultrasound using Monte Carlo Dropout and Deep Ensembles for Epistemic Uncertainty Estimation}

\titlerunning{Towards Trustworthy Breast Tumor Segmentation in Ultrasound}
%
    
\author{Toufiq Musah\inst{1, 2} \and
Chinasa Kalaiwo\inst{3} \and
Maimoona Akram\inst{4} \and 
Ubaida Napari Abdulai \inst{1} \and
Maruf Adewole\inst{5} \and
Farouk Dako\inst{7} \and
Adaobi Chiazor Emegoakor \inst{8} \and
Udunna C. Anazodo \inst{5,6} \and
Prince Ebenezer Adjei \inst{1, 2} \and
Confidence Raymond\inst{6} 
}
\authorrunning{T. Musah et. al}

\institute{
Department of Computer Engineering, Kwame Nkrumah University of Science and Technology, Kumasi, Ghana.\\ \and
Global Health and Infectious Disease Group, Kumasi Centre for Collaborative Research in Tropical Medicine, Kumasi, Ghana.\\ \and
Department of Radiology, National Hospital Abuja, Abuja, Nigeria. \\ \and
Computer Science Department, FAST National University of Computer and Emerging Sciences, Lahore, Pakistan \\ \and
Medical Artificial Intelligence Lab, Lagos, Nigeria.\\\and
Department of Biomedical Engineering, McGill University, Montreal, Canada\\\and
Perelman School of Medicine, University of Pennsylvania, 3400 Spruce Street, Philadelphia, PA 19104 \\\and
Nnamdi Azikiwe University Teaching Hospital, Nnewi, Nigeria \\
\email{toufiqmusah32@gmail.com}
}
%

\maketitle              
\begin{abstract}

Automated segmentation of BUS images is important for precise lesion delineation and tumor characterization, but is challenged by inherent artifacts and dataset inconsistencies. In this work, we evaluate the use of a modified Residual Encoder U-Net for breast ultrasound segmentation, with a focus on uncertainty quantification. We identify and correct for data duplication in the BUSI dataset, and use a deduplicated subset for more reliable estimates of generalization performance. Epistemic uncertainty is quantified using Monte Carlo dropout, deep ensembles, and their combination. Models are benchmarked on both in-distribution and out-of-distribution datasets to demonstrate how they generalize to unseen cross-domain data. Our approach achieves state-of-the-art segmentation accuracy on the Breast-Lesion-USG dataset with in-distribution validation, and provides calibrated uncertainty estimates that effectively signal regions of low model confidence. Performance declines and increased uncertainty observed in out-of-distribution evaluation highlight the persistent challenge of domain shift in medical imaging, and the importance of integrated uncertainty modeling for trustworthy clinical deployment. \footnote{Code available at: \url{https://github.com/toufiqmusah/caladan-mama-mia.git}} 

\keywords{Breast Ultrasound Segmentation \and Uncertainty Estimation \and Out-of-Distribution Data \and Deep Learning}
\end{abstract}

\section{Introduction}
Breast tumors are masses resulting from abnormal cellular proliferation within breast tissues, encompassing a broad range of pathologies, the most clinically significant of which is breast cancer. Breast cancer remains highly prevalent, and was reported as the most common cancer among females in 157 out of 185 countries \cite{who_1}, resulting in approximately 670,000 deaths in 2021, with projections indicating a constant increase in cases past 2050, especially impacting low- and middle-income regions of the world \cite{global}. Early detection and accurate diagnosis are fundamental strategies for improving survival outcomes \cite{ginsburg_3, malakouti_4}.

Multiple medical imaging techniques are employed in the detection and diagnosis of breast cancers, including mammography, breast ultrasonography (BUS), and magnetic resonance imaging (MRI). Breast ultrasonography serves as an essential complement to mammography, as it is particularly valuable for early scanning, follow-ups, and treatment monitoring \cite{iacob_7, pan_9}. It offers several practical advantages, including real-time imaging without exposure to ionizing radiation, suitability for repeated examinations, and particular effectiveness in imaging dense breast tissue commonly found in younger populations. It plays a central role in clinical workflows, especially in low- and middle-income settings where mammography or MRI may be inaccessible \cite{iacob_7}.

Accurate segmentation aids in reporting tumor features with BI-RADS \cite{BIRADs}, and clinical decision-making by improving lesion  characterization, radiation therapy planning, response monitoring, and surgical preparation \cite{rad-therapy, evolution, management}. Though ultrasound-based segmentation faces notable challenges due to inherent imaging artifacts, including low contrast, speckle noise, blurred lesion boundaries, and significant operator dependence. The diverse morphological presentation of breast tumors and limitations arising from inadequate and imbalanced datasets further complicate downstream tasks including segmentation \cite{huang-survey}.

In this body of work, we explore the use of deep learning methods in the segmentation of breast ultrasound images, and further estimate the uncertainty in model predictions using various combinations of epistemic methods. The widely used Breast Ultrasound Images (BUSI) dataset is shown to have unreliable segmentation performance due to data duplication and inconsistent annotations across the same subject images, resulting in data leakage between training and validation sets. Epistemic uncertainty is estimated via Bayesian inference approximation using Monte Carlo dropout, followed by deep ensembling. We also experiment with a combined Monte Carlo dropout–deep ensembling approach. These methods are evaluated on an out-of-distribution test set to emulate real-world deployment scenarios.



\section{Related Works}
The paradigm shifted with encoder–decoder deep neural network architectures such as U-Net \cite{Olaf} and its improved variants, including UNet++ \cite{Zongwei} and Attention‑UNet \cite{Ozan}, which significantly advanced segmentation accuracy and robustness in breast ultrasound segmentation \cite{ajay, anari_6_method, alblwi2025d}. Recent innovations include AAU‑Net \cite{chen}, HCTNet \cite{qiqi}, and LightBTSeg \cite{Guo}. Despite these advances, most methods focus on in-distribution data without robust cross-domain validation.

Epistemic uncertainty can be approximated via MC dropout \cite{gal2016dropout} or deep ensembles \cite{lakshminarayanan}. Combining these approaches produces well-calibrated uncertainty maps aligned with areas of anatomical ambiguity \cite{kurz}. In a study by Marisa et al. \cite{marisa-miccai}, epistemic uncertainty was quantified in the classification of breast tumor sub-types, exploring approximated Bayesian inference in MC dropout, and deep ensembles. These models produce coherent uncertainty estimates across anatomical regions or entire structures, enabling more meaningful confidence assessment along boundaries and complex areas. Such structured uncertainty frameworks improve interpretability, and can better support clinical decision-making where trustworthiness is necessary.

\section{Methods}
\subsection{Dataset}
\begin{figure}
    \centering
    \includegraphics[width=1.0\textwidth]{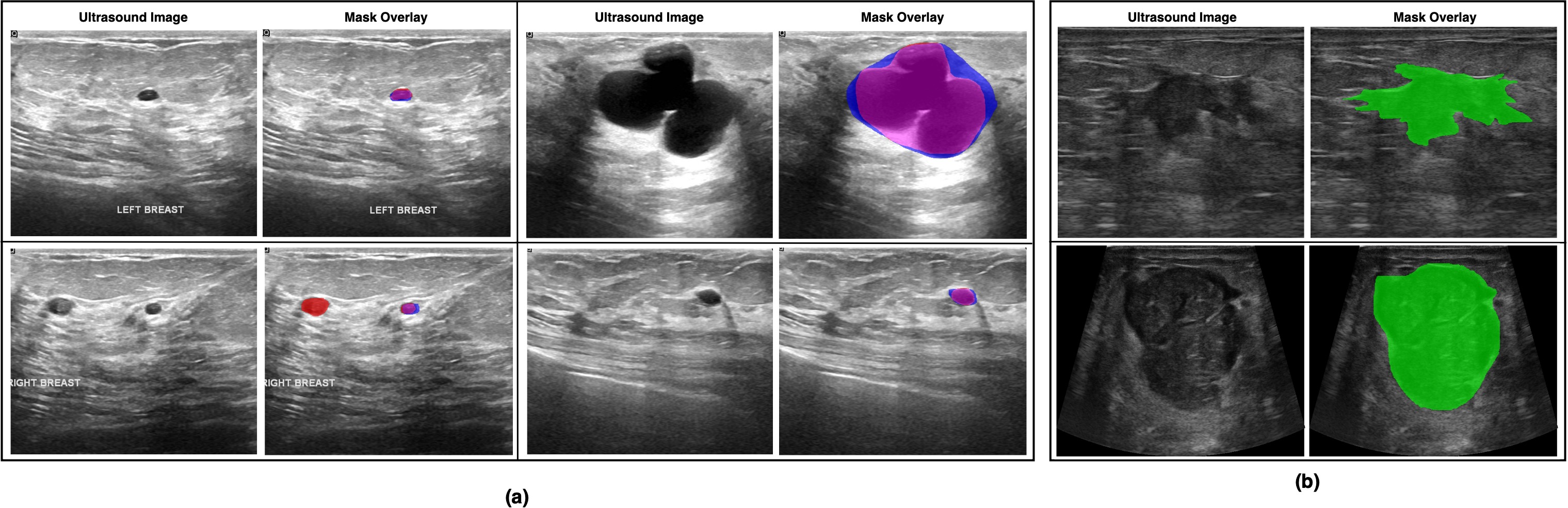}
    \caption{(a) Sample cases with duplicate masks from the training and validation sets. (\textit{Annotator-1} - Red | \textit{Annotator-2} - Blue | Overlap - Magenta) 
    (b) Sample cases from the out-of-distribution testing dataset}
    \label{fig:dataset}
\end{figure}

This study utilizes two datasets; the \textbf{Breast UltraSound Images (BUSI)} (Fig. \ref{fig:dataset} (a)) dataset \cite{BUSI} for model training and validation, and the \textbf{Breast-Lesions-USG} dataset \cite{Breast-Lesions-USG} (Fig. \ref{fig:dataset} (b)) for out-of-distribution testing and uncertainty quantification. This is to enable both in-distribution performance assessment and evaluation of model generalizability. The original BUSI dataset contained several discrepancies identified by \cite{Re-BUSI}, such as duplicated images, and the inadvertent inclusion of non-breast images (maxilla ultrasound scans), which were not explicitly stated in the dataset publication \cite{BUSI}. We further note that the duplicated sets were of varying annotations, which led us to systematically deduplicate the dataset in three distinct ways:

\begin{enumerate}
    \item \textbf{BUSI-A1:} Removed the first occurrence of each duplicate pair.
    \item \textbf{BUSI-A2:} Removed the second occurrence of each duplicate pair.
    \item \textbf{BUSI-A3:} Retained the duplicate deemed most accurate by a radiologist.
\end{enumerate}

\subsection{Modelling}
We employ a modified Residual Encoder U-Net with dropout layers, trained with the nnUNet framework \cite{isensee2024nnu}. It follows an identical setup as described in previous work \cite{musah2024automated} including 8 encoder stages and 7 decoder stages with increasing feature sizes per stage. A typical residual block in the modified setup comprises 6 layers; 

{\small $$Conv2D \rightarrow Dropout \rightarrow InstanceNorm \rightarrow LeakyReLU \rightarrow Conv2D \rightarrow InstanceNorm$$}

The models were trained with deep supervision and optimized using stochastic gradient descent with a batch dice loss. We used a patch size of $512 \times 512$ for the input of 2D breast ultrasound images, and a batch size of 13. By default, we train all folds for 250 epochs, and further train \textit{BUSI-A3} for another 750 epochs before applying it on the test dataset for out-of-distribution evaluation.

\subsection{Uncertainty Estimation}

We quantify uncertainty using three complementary methods: Monte Carlo (MC) dropout, Deep Ensembles, and a combined Deep Ensemble-MC dropout approach.

\paragraph{MC Dropout} Estimates epistemic uncertainty via multiple stochastic forward passes with dropout active at inference \cite{gal2016dropout}. For input \( x \), predictions are averaged as:

$$
p(y|x) \approx \frac{1}{T}\sum_{t=1}^T f_{\theta_t}(x),
$$

where \( f_{\theta_t}(x) \) is the prediction with dropped weights at iteration \( t \). Uncertainty is the variance across these predictions.

\paragraph{Deep Ensembles} Aggregate predictions from \( K \) independently trained models \cite{lakshminarayanan}:

$$
p(y|x) \approx \frac{1}{K}\sum_{k=1}^K f_{\theta^{(k)}}(x).
$$

Variability captures uncertainty from data splits and initialization.

\paragraph{Combined Approach} Each ensemble member performs \( T \) stochastic passes:

$$
p(y|x) \approx \frac{1}{K}\sum_{k=1}^K \frac{1}{T}\sum_{t=1}^T f_{\theta_t^{(k)}}(x).
$$

This jointly captures intra- and inter-model epistemic uncertainty. Aleatoric uncertainty is not modeled.

\subsubsection{Uncertainty Evaluation}

To quantify epistemic uncertainty at the pixel level, we evaluate on the following metrics: 

\paragraph{Predictive Entropy} For pixel \((i,j)\), mean predicted probability over \( T \) stochastic forward passes is

$$
\bar{p}_{ij} = \frac{1}{T}\sum_{t=1}^T \hat{p}_{ij}^{(t)},
\quad
\mathcal{H}(\bar{p}_{ij}) = -\bar{p}_{ij}\log \bar{p}_{ij} - (1-\bar{p}_{ij})\log(1-\bar{p}_{ij}).
$$

\paragraph{Mutual Information} Epistemic uncertainty is

$$
\mathcal{I}(y,\theta|x_{ij}) = \mathcal{H}(\bar{p}_{ij}) - \frac{1}{T}\sum_{t=1}^T \mathcal{H}(\hat{p}_{ij}^{(t)}).
$$

\paragraph{Expected Calibration Error (ECE)} Measures confidence-accuracy alignment \cite{guo-on,kull}. Pixels are binned by confidence \(\max(\bar{p}_{ij}, 1-\bar{p}_{ij})\). ECE is
$$
\text{ECE} = \sum_{m=1}^M \frac{|B_m|}{N} \left| \text{acc}(B_m) - \text{conf}(B_m) \right|,
$$

where \( |B_m| \) is bin size, \( N \) total pixels, and \(\text{acc}, \text{conf}\) are accuracy and confidence per bin. Lower ECE indicates better calibration. We use 30 bins over 83 million pixels, including 5.8 million foreground pixels.
\section{Results and Discussion}
\begin{figure}[ht]
    \centering
    \includegraphics[width=1.0\textwidth]{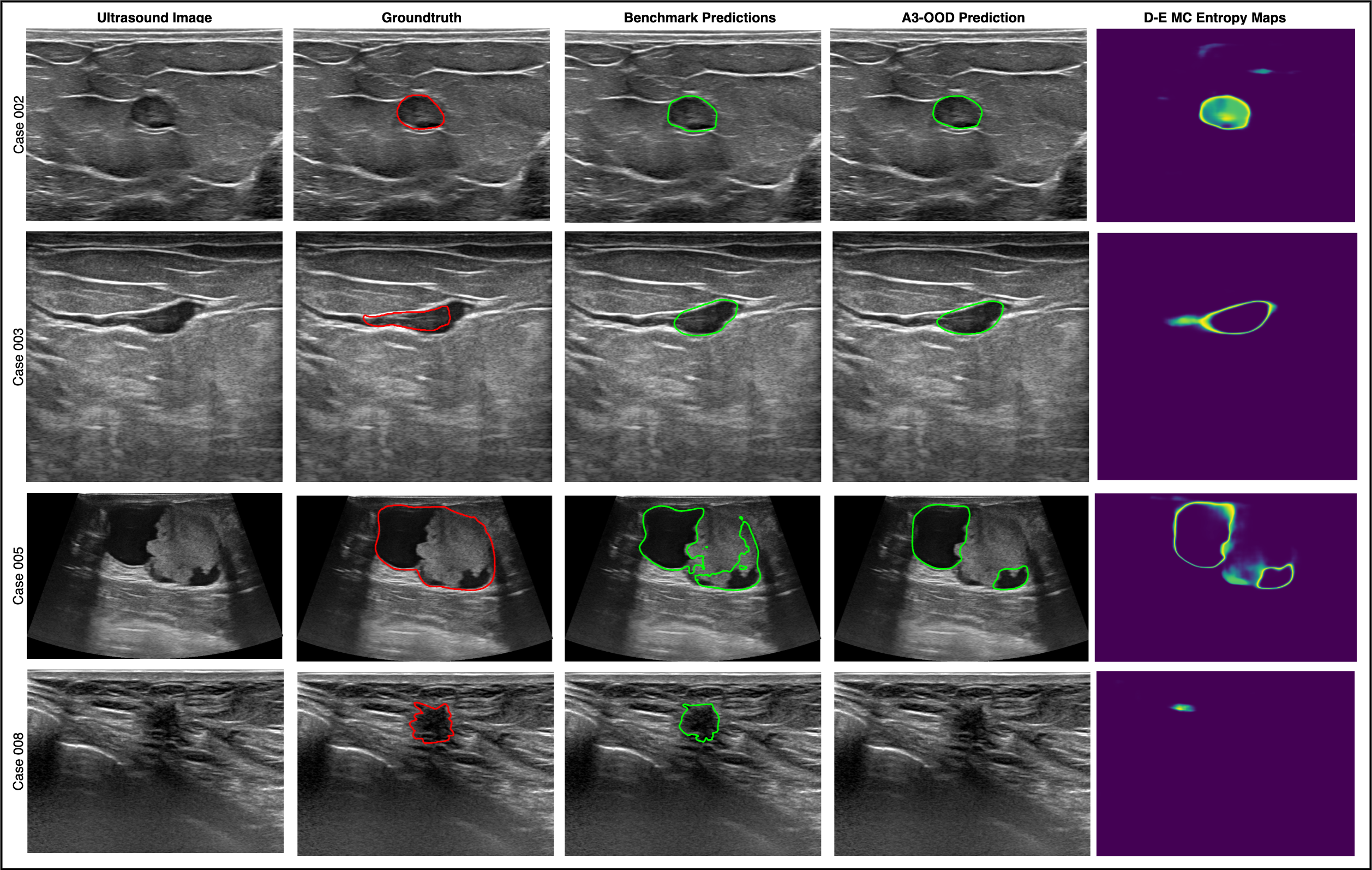}
    \caption{Qualitative examples of segmentation and uncertainty entropy maps on the Breast-Lesion-USG dataset. Columns show the original ultrasound image, ground truth annotation (red), benchmark (in-domain) and A3-OOD (out-of-distribution) predictions  (green), and the corresponding deep ensemble Monte Carlo (D-E MC) entropy map. Higher entropy (yellow) highlights regions of increased model uncertainty.}
    \label{fig:results}
\end{figure}

\subsection{Segmentation Performance}
We conducted 5-fold cross-validation on four variations of the BUSI dataset: \textit{BUSI-Full}, which includes the complete dataset with duplicates, resulting in overlapping cases between training and validation sets, and \textit{BUSI-A1, A2 and A3}, curated subsets containing only unique annotated cases.

\begin{table}[ht]
\centering
\caption{5-Fold Cross-Validation Dice Scores for BUSI Datasets}
\label{tab:busi-dice-scores}
\renewcommand{\arraystretch}{1} 
\setlength{\tabcolsep}{10pt}      
\footnotesize
\begin{tabular}{lcccc}
\toprule
\textbf{Fold} & \textbf{BUSI-Full} & \textbf{BUSI-A1} & \textbf{BUSI-A2} & \textbf{BUSI-A3} \\
\midrule
Fold 0        & 0.7478             & 0.7048                 & 0.6927                 & 0.7732 \\
Fold 1        & 0.7147             & 0.7092                 & 0.7366                 & 0.6657 \\
Fold 2        & 0.7769             & 0.6815                 & 0.7324                 & 0.7084 \\
Fold 3        & 0.7667             & 0.7512                 & 0.7260                 & 0.7670  \\
Fold 4        & 0.7509             & 0.7253                 & 0.7016                 & 0.6911  \\
\midrule
\textbf{Average} & \textbf{0.7514} & \textbf{0.7144}        & \textbf{0.7179}        & \textbf{0.7211} \\
\bottomrule
\end{tabular}
\end{table}

Table~\ref{tab:busi-dice-scores} presents the Dice scores across folds. The \textit{BUSI-Full} dataset achieved a higher average Dice score of 0.7512, compared to 0.7144, 0.7179, and 0.7211 for \textit{A1}, \textit{A2}, and \textit{A3}, respectively. This difference may be attributed to data leakage between the training and validation sets in the original dataset. The performance on \textit{BUSI-A1, A2 and A3} are therefore considered more indicative of true generalization. Lower scores are expected when data leakage and redundancy are corrected, as evaluation is no longer artificially inflated by overlaps between training and validation sets.

\begin{table}[ht]
\centering
\caption{
  Comparison of State-of-the-Art methods on Breast-Lesion-USG vs. our methods using in-distribution validation (\textbf{Benchmark}) and OOD validation (\textit{A3-OOD}).
}
\label{tab:benchmark-comparison}
\renewcommand{\arraystretch}{1}
\begin{tabular}{lcc}
\toprule
\textbf{Model} & \textbf{Dice} & \textbf{IoU} \\
\midrule
ResUNet                & 0.4563 & 0.3444 \\
UNet++                 & 0.3734 & 0.2564 \\
Attention-UNet         & 0.4764 & 0.3000 \\
SwinUNet               & 0.4436 & 0.3331 \\
D-DDPM \cite{alblwi2025d}  & 0.7104 & 0.6140 \\
\textit{Ours$_{Benchmark}$}      & \textbf{0.7726} & \textbf{0.6801} \\ \midrule
\textit{Ours$_{A3-OOD}$}                 & \textit{0.4855 }   & \textit{0.4309 }   \\
\bottomrule
\end{tabular}
\end{table}

We further benchmarked our method on the Breast-Lesion-USG dataset, training for 250 epochs using 5-fold cross-validation to provide a measure of segmentation performance when the model is trained and validated on the same data distribution. We achieved an average Dice score of 0.7726 $\pm$ 0.0212 across folds.

To evaluate the out-of-distribution performance of the model, we selected the \textit{BUSI-A3} subset, which we recommend as the most representative, as it was deduplicated by a trained radiologist. We compare this performance against other methods reported in \cite{alblwi2025d}, evaluating on Dice scores, and Intersection over Union (IoU) summarized in Table~\ref{tab:benchmark-comparison}.

\subsection{Uncertainty Estimation Results}
\begin{table}[ht]
\centering
\setlength{\tabcolsep}{3pt} 
\caption{Summary of Uncertainty and Calibration Metrics Across Methods}
\label{tab:uncertainty-summary}
\footnotesize
\begin{tabular}{lcccc}
\toprule
\textbf{Method} & \textbf{Entropy(\textcolor{gray}{$\downarrow$})} & \textbf{MI(\textcolor{gray}{$\downarrow$})} & \textbf{ECE(\textcolor{gray}{$\downarrow$})} & \textbf{Pixel-wise Acc.(\textcolor{gray}{$\uparrow$})} \\
\midrule
Monte Carlo (MC) Dropout                & 0.009  & 0.002  & 0.0367 & 0.9595 \\
Deep-Ensemble (D-E)           & 0.021  & 0.013  & 0.0300 & 0.9607 \\
D-E MC Dropout       & 0.031  & 0.019  & 0.0303 & 0.9604 \\
\bottomrule
\end{tabular}
\end{table}

We evaluated model uncertainty using three strategies: Monte Carlo (MC) dropout, deep ensembles, and a combined deep ensemble MC dropout approach. All analyses were conducted on the 256 cases of the Breast-Lesions-USG dataset \cite{Breast-Lesions-USG}, comprising a total of 83,099,921 analyzed pixels, summarized in Table~\ref{tab:uncertainty-summary}.

\paragraph{Monte Carlo Dropout.}
MC dropout yielded a mean predictive entropy (total uncertainty) of $0.009$ (range: $[0.000, 0.693]$; average standard deviation within cases: $0.046$), and a mean epistemic uncertainty (mutual information) of $0.002$ (range: $[0.000, 0.488]$; average within-case standard deviation: $0.010$) on 10 stochastic forward passes. The median entropy and mutual information across cases were both near zero, indicating that most pixels were predicted with high confidence.

\paragraph{Deep Ensemble.}
Using a deep ensemble, the model exhibited a mean predictive entropy of $0.021$ (range: $[0.000, 0.693]$; average standard deviation: $0.076$) and a mean epistemic uncertainty of $0.013$ (range: $[0.000, 0.673]$; average standard deviation: $0.050$).

\paragraph{Deep Ensemble Monte Carlo Dropout.}
For the combined approach (5 ensemble members $\times$ 3 MC dropout samples each; $15$ samples per case), the mean predictive entropy increased to $0.031$ bits, and mean mutual information to $0.019$ bits.

Notably, when evaluating \textit{Ours$_{A3-OOD}$} (out-of-distribution), we observed a substantial drop in Dice and IoU scores relative to in-domain performance (\textit{Benchmark}) (see Table~\ref{tab:benchmark-comparison}). This decline in segmentation accuracy was accompanied by increased predictive entropy and mutual information values, reflecting the model’s heightened epistemic uncertainty when faced with unfamiliar inputs.

\section{Discussion and Conclusion}
Our benchmarking on the Breast-Lesion-USG dataset using standard in-domain cross-validation showed that our method achieves state-of-the-art Dice and IoU scores, outperforming models including ResUNet, UNet++, SwinUNet, and D-DDPM \cite{alblwi2025d}. However, training on the strictly deduplicated BUSI-A3 subset and testing out-of-distribution on Breast-Lesion-USG results in a significant drop in segmentation accuracy. This shows the persistent challenge of domain shift and the need for models that generalize reliably across diverse datasets \cite{huang-survey}.

All uncertainty quantification methods and their combination, yielded low average predictive entropy and mutual information, indicating high confidence in the prediction of most pixels. D-E and combined D-E MC showed better Expected Calibration Error (ECE) \cite{guo-on, kull} and pixel-wise accuracy than MC alone. On out-of-distribution data, higher uncertainty values corresponded with decreased accuracy, with entropy and mutual information effectively highlighting unreliable prediction regions. Clinically, these findings emphasize the importance of robust dataset preparation to avoid optimistic generalization estimates, and highlight uncertainty quantification as an important safeguard in decision support, enabling practitioners to recognize and manage predictions under uncertainty or domain shift.

\section{Limitations and Future Work}
While entropy maps may provide intuitive qualitative uncertainty visualization, quantitative calibration via ECE depends on binning schemes that can be sensitive and may not generalize across different data distributions. Pixel-wise accuracy tends to be inflated because of the imbalance between foreground and background pixels. Future work should employ segmentation-aware calibration methods \cite{zeevi} to obtain more realistic estimates. Further, MC Dropout and Deep Ensembles increase inference time by 10 to 25 times as compared to single forward passes, posing challenges for real-time clinical applications.

In datasets like BUSI-Full with multiple annotator segmentations, human uncertainty calibration could align model confidence with clinical opinion variability rather than relying on majority votes. Collecting such datasets with intentional design is valuable. Although our methods advance ultrasound breast lesion segmentation, they expose limitations in cross-domain deployment. Future work should focus on adaptive models to bridge generalization gaps and refine uncertainty estimation for safer and more transparent clinical integration.

\section*{Acknowledgments}
This work was supported by the Lacuna Fund on Sexual, Reproductive and Maternal Health and Rights (SRMHR) for the African Breast imaging dataset for equitable cancer care (ABreast data) Project and completed as part of the 2024 Precision Cancer Care in Africa (PRECISE) Symposium Hackathon in collaboration with the University of Pennsylvania Center for Global and Population Health Research in Radiology, the Medical Artificial Intelligence Laboratory (MAI Lab), the National Institute for Cancer Research and Treatment (NICRAT) Nigeria, Ernest Cooke Ultrasound Research and Education Institute Uganda, Consortium for Advancement of MRI Education and Research in Africa (CAMERA).

\bibliographystyle{unsrt}
\bibliography{references}

\end{document}